\documentstyle[twocolumn,aps]{revtex}
\begin{document}
\draft
\title{\large {\bf
IDENTICAL BANDS IN SUPERDEFORMED NUCLEI:\\
A RELATIVISTIC DESCRIPTION\cite{AAA}}}
\author{J. K\"onig and P. Ring}
\address{Physik-Department der Technischen Universit\"at M\"unchen,\\
D-8046 Garching, FRG}
\date{\today}
\maketitle
\begin{abstract}
Relativistic Mean Field Theory in the rotating frame is
used to describe superdeformed nuclei. Nuclear currents and
the resulting spatial components of the vector meson fields
are fully taken into account. Identical bands in
neighboring Rare Earth nuclei are investigated and
excellent agreement with recent experimental data is
observed.
\end{abstract}

\narrowtext

Since the experimental discovery of superdeformed bands in
rapidly rotating nuclei many unexpected features of these
highly excited configurations have been observed (for a
recent review see Ref. \cite{JK.91}). One of the most
striking properties is the existence of so-called {\it
identical bands} or {\it twin bands}, i.e. nearly identical
transition energies $E_\gamma$ of the emitted
$\gamma$-radiation in bands belonging to neighboring nuclei
with different mass numbers. In a considerable number of
nuclei in the Dy-region as well as in the Hg region one has
found differences in $E_\gamma$ of only 1-3 keV, i.e. there
exists sequences of bands in neighboring nuclei, which are
virtually identical, $\Delta E_\gamma/E_\gamma\sim 10^{-3}$. Since
these transition energies are directly related to the
corresponding dynamical moments of inertia obeying on the
average a simple $A^{5/3}$ dependence, one would have
expected changes of one order of magnitude larger.

Several groups have tried to understand this phenomenon by
means of conventional investigations within the framework of
the semi-phenomenological Strutinski method in connection
with  a rotating Nilsson or Saxon-Woods
potential\cite{WD.91,Ra.90}. It has been pointed out, that
the different single particle orbits can give rather
different contributions to the moment of inertia.  This is
most clearly seen in the simple oscillator model, where
orbits with oscillator quanta along the rotational axis
have vanishing angular momentum operator matrix elements.
Other groups\cite{NTF.90} realized that a new coupling scheme
in nuclei exists, the so-called pseudo-spin scheme,
which to a very large extent decouples the
pseudo-orbital motion from the pseudo-spin degrees of
freedom and favors the strong coupling limit. In all these
investigations, however, polarization effects, which are
expected to produce much larger changes of the moments of
inertia than those observed in identical bands, are either
neglected completely\cite{NTF.90} or taken into account
only partially by minimizing the rotating energy surface
with respect to a few deformation parameters.

We therefore feel that it is very important to carry out
fully self-consistent microscopic calculations, where all
the degrees of freedom are taken into account. Such
calculations are not simple, but they are nowadays
feasible. A first investigation of this type using the
density dependent Skyrme III force has very recently been
carried out in the Hg region\cite{CHB.92}. It has been
found that full self-consistency has indeed a considerable
influence on the details of the moment of inertia in
neighboring nuclei. There are, however, many important
questions still open. In these nuclei identical bands occur
in a spin region, where pairing plays an important role.
Since particle-particle correlations are not very well
described within the Skyrme scheme, it is not at all clear,
to what extent the present deviations of the theoretical
from the experimental results can be understood by such
deficiencies.

The present investigation is therefore devoted to the
Dy-region, where superdeformed bands are observed up to
very high angular momenta. At these very large rotational
frequencies the Coriolis-anti-pairing effect reduces these
correlations considerably, such that they have little
influence on the moment of inertia. We use a relativistic
field theory which includes $\sigma$-, $\omega$- and
$\rho$-mesons as well as the electromagnetic field. In
addition we consider a nonlinear self-coupling of the
$\sigma$-field. We use the parameter set NL1, which has
been adjusted\cite{RRM.86} to nuclear matter and a few
spherical nuclei. This parameter set has turned out to be
very successful for the description of many groundstate
properties over the entire periodic table\cite{GRT.90}. In
particular one has found excellent agreement with ground
state deformations in open shell nuclei.  The rotation is
treated within the cranking approach, in accordance with
the concept of a mean field description. This leads us to a
relativistic self-consistent cranking theory (RSCC) as
developed in Ref. \cite{KR.89}.

In the rotating frame time reversal invariance is broken.
This lead to nucleonic currents in the interior of the
nucleus, which form the source of magnetic potentials in
the Dirac equation ({\it nuclear magnetism}). In this way
the charge current ${\bf j}_c$ is the source of the normal
magnetic potential ${\bf A}$, the isoscalar baryon current
${\bf j}_B$ is the source of the spatial components
{\boldmath $\omega$}
of the $\omega$-mesons and the isovector baryon
current ${\bf j}_3$ is the source of the spatial component
{\boldmath $\rho$}$_3$ of the $\rho$-mesons. In contrast to the
maxwellian magnetic field $\bf A$ having a small
electromagnetic coupling, the large coupling constants of
the strong interaction causes the fields {\boldmath $\omega$} and
{\boldmath $\rho$} to be important in all cases, where they are not
forbidden by symmetries, such as time reversal. They have a
strong influence on the magnetic moments\cite{HR.88} in odd
mass nuclei, where time reversal is broken by the odd
particle, as well as the moment of inertia in rotating
nuclei, where time reversal is broken by the Coriolis
field. In an early investigation of rapidly rotating
superdeformed nuclei within the framework of cranked
relativistic mean field theory\cite{KR.90} these components
were not taken into account for reasons of numerical
simplicity. Strong deviations from the experimentally
observed moments of inertia were found.  Only the size
of the quadruple deformation was reproduced properly.
Semiclassical corrections turned out to be large, but
could not reproduce the proper experimental values of the
moment of inertia.

In this investigation nuclear magnetism is taken fully into
account in a self-consistent way. Starting from the
Lagrangian
\begin{eqnarray}
{\cal L}&=&
\bar\psi\left(\rlap{/}p-g_\omega\rlap{/}\omega
+g_\rho\rlap{/}\vec\rho\vec\tau-\frac{1}{2}e(1+\tau_3)\rlap{\,/}A
-g_\sigma\sigma-M_N\right)\psi\nonumber\\
&&+\frac{1}{2}\partial_\mu\sigma\partial^\mu\sigma-U(\sigma)
-\frac{1}{4}\Omega_{\mu\nu}\Omega^{\mu\nu}+
\frac{1}{2} m^2_\omega\omega_\mu\omega^\mu\\
&&-\frac{1}{4}\vec R_{\mu\nu}\vec R^{\mu\nu}+
\frac{1}{2} m^2_\rho\vec\rho_\mu\vec\rho^\mu
-\frac{1}{4}F_{\mu\nu}F^{\mu\nu}
\nonumber
\end{eqnarray}
where $M_N$ is the bare nucleon mass and $\psi$ is its
Dirac spinor. We have in addition the scalar meson
($\sigma$), isoscalar vector mesons ($\omega^\mu$),
isovector vector mesons ($\vec\rho^\mu$) and the photons
$A^\mu$, with the masses $m_\sigma$, $m_\omega$ and
$m_\rho$ and the coupling constants $g_\sigma$, $g_\omega$,
$g_\rho$. For simplicity in the
following equations we neglect the $\rho$-meson and
the photon. In the calculations these contributions are,
however, taken into account however. The field tensors for the
vector mesons are given as
\begin{equation}
\Omega_{\mu\nu}~=~\partial_\mu\omega_\nu-\partial_\nu\omega_\mu,
\end{equation}
For a realistic description of
nuclear properties a nonlinear self-coupling for the scalar
mesons has turned out to be crucial\cite{BB.77}:
\begin{equation}
U(\sigma)~=~\frac{1}{2} m^2_\sigma \sigma^2_{}
{}~+~\frac{g_2}{3}\sigma^3_{}~+~\frac{g_3}{4}\sigma^4_{}
\end{equation}
Starting from this Lagrangian and transforming to a frame
rotating with a uniform velocity $\Omega_x$ around the
$x$-axis perpendicular to the symmetry axis of the deformed
nucleus in its ground state, we obtain the classical
equations of motion
\begin{equation}
\{ {\mbox{\boldmath $\alpha$}}
({\bf p}+g_\omega{\mbox{\boldmath $\omega$}})+g_\omega\omega_0
+\beta(M+g_\sigma\sigma)-\Omega_x J_x\}
\psi_\alpha=\epsilon_\alpha\psi_\alpha
\end{equation}
for the nucleon spinors and
\begin{mathletters}
\begin{eqnarray}
\left\{-\Delta+(\Omega_xL_x)^2\right\}\sigma~+~U'(\sigma)&=&
-g_\sigma\rho_s
\\
\left\{-\Delta+(\Omega_xL_x)^2~+~m_\omega^2\right\}\omega^0&=&
g_\omega\rho_v
\\
\left\{-\Delta+(\Omega_xJ_x)^2~+~m_\omega^2\right\}
{\mbox{\boldmath $\omega$}}^{~}&=&
g_\omega{\bf j}
\label{mesonmotion}
\end{eqnarray}
\end{mathletters}
where $J_x=L_x+S_x$ and the spin operator $S_x$ is a $4\times
4$-matrix
for the spinor fields with spin $\frac{1}{2}$ and a $3\times 3$-matrix
for
vector fields with spin 1. For details see Ref. \cite{KR.89}.

These equations are solved self-consistently by expanding
the Dirac spinors as well as the meson fields in terms of
eigenfunctions of a deformed oscillator, as discussed in
details in Refs. \cite{KR.89,GRT.90}. Up to $N_F=13$ major
oscillator shells where taken into account for the large
components of the Fermions fields and up to $N_B=13$ shells
for the meson fields. Because of the large number of
configurations in this spaces high-lying orbitals with a
deformed oscillator energy larger than
$10.3\times\hbar\omega^0$ for the Fermions and larger than
$10.5\times\hbar\omega^0$ for the Bosons have been
neglected. In general one has to allow for complex
expansion coefficients in this basis. However, assuming
mirror symmetry at the three planes ($x,y$), ($x,z$), and
($y,z$) for the densities and rotational symmetry for the
currents it is possible to restrict oneself to real
coefficients. Since, in the rotating frame, their is no
a priori reason for these symmetries, a complex
code was used to show that, even with initial conditions
strongly violating this symmetries, we find after many
iterations final self-consistent solutions which obey these
symmetries.  All the following calculations have therefore
been carried out with the real code only.

In Fig. 1 we show the static  and the dynamic moment of
inertia for the lowest superdeformed band in the nucleus
$^{152}$Dy as a function of the angular momentum. It is
clearly seen that a calculation without nuclear magnetism,
i.e. without the spatial contributions of the vector
meson-fields, which is in good agreement with the
experimental quadrupole moments (see Ref. \cite{KR.90})
produces much too small moments of inertia. A semiclassical
correction where these contributions, derived in Thomas
Fermi approximation using a rigid rotor current, are taken
into account in first order perturbation theory,
overemphasized the moments of inertia by roughly 10
\%. Only if one takes these contributions into account in a
fully self-consistent way, is perfect agreement with
experimental data achieved. In the region of small
angular momenta one still observes very small deviations,
which could possibly be understood as the influence of remaining
pairing correlations in this region of intermediate spins.

We also find that nuclear magnetism practically has no
influence on the shape of the nucleus. The mass quadrupole
moments decrease in the spin range from 20 to 60 $\hbar$
only very little, running from 4350 to 4260 fm$^2$ and the
corresponding hexadecupole moments change for the same
region from 20600 to 19600 fm$^4$. The changes induced by
nuclear magnetism are of the order of a few per mille.
The average charge quadrupole moment if found to be 18.6 eb,
which is in good agreement with a value of 18 eb obtained in a
non-relativistic calculation\cite{RA.86}, and the
experimental value of 19 eb \cite{BCF.87}
 From the quadrupole moments we can derive the Hill-Wheeler
parameters $\beta=0.72$ and $\gamma=0.7^\circ$ for the
quadrupole deformations, which corresponds closely
to a nearly prolate deformed nucleus with an
axis ratio of 1:1.9, close to the standard value 1:2 of the
harmonic oscillator model.

Let us now investigate the problem of identical bands. For
this purpose we calculate, in a self-consistent way, bands in
the neighboring nucleus $^{151}$Tb by removing one proton
from the $^{152}$Dy core. In Fig. 2 we show the single
particle spectrum for protons in the rotating potential
formed by the $^{152}$Dy core. The large gap at Z = 66 is
clearly recognized. Taking particles out of the orbits
directly below this gap, we can produce different bands in
the nucleus $^{151}$Tb. They have the quantum numbers (PS)
of parity (P) and signature (S), namely ($+-$) for the
dashed line, ($++$) for the full line, ($-+$) for the
dotted line and ($-+$) for the dashed dotted line.

The proton hole induces a polarization of the
$^{152}$Dy-core, which has two effects: it leads to changes
of deformation and in addition to changes in the
current distribution. In Table \ref{T1} we show the values obtained
after solving in a fully self-consistent fashion the
relativistic mean field equations for the odd system in the
four lowest configurations. We show the values for several
observables for the lowest superdeformed band in
$^{152}$Dy. The relative changes with respect to this
reference band in the four bands of $^{151}$Tb are given in
per mille.  According to the simple $A^{5/3}$-rule we
expect changes in the moment of inertia by $\approx$ 11 per
mille. The calculated values for the moments of inertia
${\cal J}^{1}$ and ${\cal J}^{2}$ for the band with the
quantum numbers ($-+$), which we we shall in the following
call the {\it identical band}, are, however, at least an order of
magnitude smaller. This is by no means trivial, because we
find considerably larger changes in the quadrupole moments
and in the rigid body moments of inertia. In fact in most
of the other bands the changes are also much larger.

In order to have a direct comparison with the experiment we
show in Fig. 3 the differences $\Delta E_\gamma=
E_\gamma($Tb$)-E_\gamma($Dy) between the transition
energies in several bands in the nucleus $^{151}$Tb and in
the lowest superdeformed band in $^{152}$Dy. The agreement
with the experimental value is excellent for the band with the
quantum numbers ($-+$), where the energy differences are of
order of 1 keV. As we see in Fig. 2, this band
correspond to a hole in the orbit with the approximate
Nilsson quantum numbers [301]$\frac{1}{2}-$. This orbit has
a very small number of oscillator quanta along the $z$-axis
(the symmetry axis), which yields nearly vanishing
contributions to the moment of inertia. We are therefore in
agreement with the qualitative argument put forward in Ref.
\cite{BBC.90}

This is, however, not the full story. In order to
investigate the very good quantitative agreement, we have
carried out two additional calculations in Fig. 4 for the
{\it identical band} band with the quantum numbers ($-+$):
First we neglected the polarization induced by the proton
hole, i.e we calculated the energy differences for wave
functions for the nucleus $^{151}$Tb obtained from the
$^{152}$Dy core by just removing one proton, without
requiring self-consistency for the odd mass configuration.
In this case we find the dashed dotted line in Fig. 4,
which is in sharp disagreement with the experimental data.
Next we took into account the polarization, but we
neglected nuclear magnetism, i.e.  the contributions of the
spatial components of the vector meson fields and find the
dashed line in Fig. 4, which is also in disagreement with
experiment.

We therefore conclude, that a very delicate cancellation
process occurs in identical bands in superdeformed nuclei.
Polarization of the quadrupole moments and of the density
alone would induce changes of the order of 5 -- 10 per
mille. Neglecting nuclear magnetism would also lead to
changes of this order of magnitude. Obviously both act in
opposite direction, such that the finial differences are
only in the order of 1 per mille. So far the precise
mechanism for this cancellation is not fully understood. It
requires definitely much more systematic investigations.
Nonetheless it seems to us a very satisfying and surprising
result, that without any free parameter, and simmply using the set
NL1 adjusted to nuclear matter and a few spherical nuclei,
long before identical bands had been identified, we can
obtain this degree of accuracy in the relatively simple
minded relativistic mean field approach. We have to
emphasize, however, that full self-consistency as well as
the inclusion of the nuclear currents are very important in
this context.


\newpage

\leftline{\Large {\bf Figure Captions}}
\parindent = 2 true cm
\begin{description}
\item[Fig. 1] (a) Static (${\cal J}^{(1)}$) and (b) dynamic
(${\cal J}^{(2)}$) moment of inertia for the lowest
superdeformed band in the nucleus $^{152}$Dy. The dashed
line corresponds to the calculation without the spatial
contributions of the vector mesons. In the dotted line such
contributions are taken into account in a semiclassical way
and the full line represents the self-consistent solution
including these contributions fully.

\item[Fig. 2] Single proton spectra in the self-consistent
rotating potential of the superdeformed band in $^{152}$Dy
as a function of the cranking frequency $\Omega_x$. For
small values of $\Omega_x$, where the two signatures are
nearly degenerate, we indicate the approximate Nilsson
quantum numbers.

\item[Fig. 3] Differences $\Delta E_\gamma$ in the
transitional energies $E_\gamma=E(I)-E(I-1)$ between the
lowest bands for each pair of quantum numbers (P,S) in the
nucleus $^{151}$Tb and the superdeformed band in $^{152}$Dy
are compared with experimental values for the excited
superdeformed band with negative parity in $^{151}$Tb.

\item[Fig. 4] Differences $\Delta E_\gamma$ for the
identical band with the quantum numbers ($-+$). The fully
self-consistent solution (full line) and solutions
neglecting nuclear magnetism (dashed line) or polarization
induced by the proton hole (dashed dotted line) are
compared with the experiment (empty diamonds).
\end{description}

\newpage
\leftline{\Large {\bf Table Caption}}
\parindent = 2 true cm
\begin{table}\centering
\caption{
Binding energy $E$,
mass quadrupole moment $Q_0$, dynamic (${\cal J}^{2}$),
static (${\cal J}^{1}$) and rigid body (${\cal J}_{rig}$)
moment of inertia at the angular momentum $I=50\hbar$
for the superdeformed band in $^{152}$Dy and relative
changes of these values (given in \%)
several bands in the neighboring nucleus $^{151}$Tb at the same
angular momentum.}
\label{T1}
\smallskip
\begin{tabular}{*{1}{l}*{7}{c}}
\\
\multicolumn{1}{c}{Band}
&\multicolumn{1}{c}{$E$ (MeV)}
&\multicolumn{1}{c}{$Q_0$ (fm$^2$)}
&\multicolumn{1}{c}{$J^{(2)}$}
&\multicolumn{1}{c}{$J^{(1)}$}
&\multicolumn{1}{c}{$J_{rig}$}\\
\\
\hline
\\
$^{152}$Dy        & -1228.358 & 4287.7 & 82.544 & 86.41 & 93.35\\
$^{151}$Tb$^*(+,+)$& 0.53 & -2.91 & -1.554 & -1.25 & -1.69 \\
$^{151}$Tb$^{~}(+,-)$& 0.51 & -3.17 & 0.654 & 0.35 & -1.80 \\
$^{151}$Tb$^*(-,+)$& 0.56 &  1.30 & -0.001 & 0.10 & -0.45\\
$^{151}$Tb$^*(-,-)$& 0.54 &  1.19 & 0.145 & 1.48 & 0.39 \\
\\
\end{tabular}
\end{table}

\end{document}